\documentclass[12pt]{article}

\usepackage{amssymb}
\usepackage[dvips]{graphicx}

\def\appendix#1{
  \addtocounter{section}{1}
  \setcounter{equation}{0}
  \renewcommand{\thesection}{\Alph{section}}
  \section*{Appendix \thesection\protect\indent \parbox[t]{11.715cm} {#1}}
  \addcontentsline{toc}{section}{Appendix \thesection\ \ \ #1}
  }

\def\l{\Big(}
\def\r{\Big)}
\def\E{{\cal E}}
\newcommand {\eq}{\begin{equation}}
\newcommand {\beq}{\begin{equation}}
\newcommand {\eeq}{\end{equation}}
\newcommand {\beqa}{\begin{eqnarray}}
\newcommand {\eeqa}{\end{eqnarray}}
\newcommand {\n}{\nonumber \\}
\newcommand {\tr}{{\rm tr\,}}

\newcommand {\ee}{\mbox{e}}

\newcommand {\dd}{\mbox{d}}

\newcommand {\del}{\partial}

\newcommand {\defeq}{\stackrel{\rm def}{=}}

\font\mybb=msbm10 at 12pt
\def\bb#1{\hbox{\mybb#1}}

\def\IC{{\bb C}}
\def\IR{{\bb R}}
\def\IZ{{\bb Z}}

\newcommand{\vl}{\ell}
\newcommand{\K}{s}
\renewcommand{\theequation}{\thesection.\arabic{equation}}
\def\rref#1{(\ref{#1})}

\begin{document}

\setlength{\oddsidemargin}{0cm}
\setlength{\baselineskip}{7mm}

\begin{titlepage}

\baselineskip=14pt

 \renewcommand{\thefootnote}{\fnsymbol{footnote}}
\begin{normalsize}
\begin{flushright}
\begin{tabular}{l}
NBI-HE-99-44\\
ITEP-TH-58/99\\
hep-th/9911041\\
\hfill{ }\\
November 1999
\end{tabular}
\end{flushright}
  \end{normalsize}

{}~~\\

\vspace*{0cm}
    \begin{Large}
       \begin{center}
{Finite $N$ Matrix Models of Noncommutative Gauge Theory}

       \end{center}
    \end{Large}
\vspace{1cm}

\begin{center}
           J. A{\sc mbj\o rn}$^{1)}$\footnote
            {
e-mail address :
ambjorn@nbi.dk},
           Y.M. M{\sc akeenko}$^{1)\,2)}$\footnote
            {
e-mail address :
makeenko@nbi.dk},
           J. N{\sc ishimura}$^{1)}$\footnote{
Permanent address : Department of Physics, Nagoya University,
Nagoya 464-8602, Japan,\\
e-mail address : nisimura@nbi.dk}
           {\sc and}
           R.J. S{\sc zabo}$^{1)}$\footnote
           {e-mail address : szabo@nbi.dk}\\
      \vspace{1cm}
        $^{1)}$ {\it The Niels Bohr Institute\\ Blegdamsvej 17, DK-2100
                 Copenhagen \O, Denmark}\\[4mm]
        $^{2)}$ {\it Institute of Theoretical and Experimental Physics}\\
               {\it B. Cheremushkinskaya 25, 117218 Moscow, Russia} \\
\end{center}

\vskip 1 cm

\hspace{5cm}

\begin{abstract}
\noindent
We describe a unitary matrix model which is constructed from discrete analogs
of the usual projective modules over the noncommutative torus and use it to
construct a lattice version of noncommutative gauge theory. The model is a
discretization of the noncommutative gauge theories that arise from toroidal
compactification of Matrix theory and it includes a recent proposal for a
non-perturbative definition of noncommutative Yang-Mills theory in terms of
twisted reduced models. The model is interpreted as a manifestly star-gauge
invariant lattice formulation of noncommutative gauge theory, which reduces to
ordinary Wilson lattice gauge theory for particular choices of parameters. It
possesses a continuum limit which maintains both finite spacetime volume and
finite noncommutativity scale. We show how the matrix model may be used for
studying the properties of noncommutative gauge theory.
\end{abstract}
\vfill
\end{titlepage}
\vfil\eject
\setcounter{page}{2}
%%%%
%\setcounter{footnote}{0}
%\section{Introduction}
%\setcounter{equation}{0}
%\renewcommand{\thefootnote}{\arabic{footnote}}
%%%%%%%%%%%%%%%
\setcounter{footnote}{0}

\baselineskip=18pt

\section{Introduction}

\setcounter{equation}{0}
\renewcommand{\thefootnote}{\arabic{footnote}}

Noncommutative gauge theory first appeared in string theory within the
framework of toroidal compactifications of Matrix theory \cite{CDS}. It was
argued that compactification on a noncommutative torus corresponds in
11-dimensional supergravity to null tori with a non-vanishing light-like
component of the background three-form tensor field. Subsequently it was
realized that the deformation to a noncommutative space can be described in
Type II string theory as the effect of turning on a constant Neveu-Schwarz
two-form tensor field $B_{\mu\nu}$ in the worldvolumes of D-branes
\cite{MD,SW}. The parameter $\theta$ which deforms the space of functions on
the worldvolume to a noncommutative algebra is related to the $B$-field
background by $\theta\sim B^{-1}$. The low-energy effective field theory for
the gauge fields living on the D-brane worldvolume 
can be described by a noncommutative gauge
theory.

A non-trivial issue concerns the renormalizability of such gauge theories,
given their unusual non-polynomial interactions. The perturbative
renormalization properties of noncommutative Yang-Mills theory have been 
studied in \cite{ren}. 
In this paper we will present a constructive definition of
noncommutative Yang-Mills theory which is the analog of the usual Wilson
lattice gauge theory \cite{Wilson}
in the commutative case. Such a model has the potential of
clarifying issues of renormalization as well as shedding light on
non-perturbative aspects of the gauge theory. A concrete definition of
noncommutative gauge theory has been proposed recently in \cite{AIIKKT}, and
further studied in \cite{IIKK,BM}, based on a large $N$ reduced model
\cite{EK}--\cite{IKKT}. In this case an ultraviolet regularization is naturally
introduced at finite $N$ and is removed in the large $N$ limit with an
appropriate fine-tuning of the gauge coupling constant. One expects the
resulting theory in the continuum limit to have three scale parameters, the
extent $L$ of the space-time, the scale $\lambda$ of noncommutativity, and the
usual gauge theoretic scale parameter $\Lambda$. However, it is found that
$\frac{L}{\lambda}$ scales as $\sqrt{N}$, which means that one is inevitably
led either to a finite $L$ with $\lambda=0$ (commutative finite space) or to a
finite $\lambda$ with $L=\infty$ (noncommutative infinite space) \cite{IIKK}.
In the following we will show that there exists a more general constructive
definition of noncommutative gauge theory which possesses a continuum limit
whereby noncommutativity is compatible with a finite volume space.

The noncommutative gauge theory that naturally arises from toroidal
compactification of Matrix Theory \cite{CDS} comes from the matrix model which
is obtained by dimensionally reducing ordinary Yang-Mills theory to a point
\cite{IKKT}. The action is
\beq
{\cal S}=-\frac1{g^2}\sum_{\mu\neq\nu}\tr\,[X_\mu,X_\nu]^2
\label{IKKT}\eeq
where $X_\mu$, $\mu=1,\dots,d$, are $N\times N$ hermitian matrices and $d$ is
the dimension of spacetime. To describe the compactification of this model on,
say, a two-torus of radii $R_1$ and $R_2$, one needs to restrict the action
(\ref{IKKT}) to those matrices $X_\mu$ that remain in the same gauge orbit
after a shift by $2\pi R_\mu\,{\bf1}_N$ in the direction $\mu$. This is
tantamount to finding configurations for which there exists unitary matrices
$\Omega_\mu$, $\mu=1,2$, which generate the quotient conditions
\cite{CDS,BFSS,cmp}
\beq
X_\mu+2\pi R_\mu\delta_{\mu\nu}\,{\bf1}_N=\Omega_\nu X_\mu\Omega_\nu^\dag \ .
\label{quotientconds}\eeq
Taking the trace of both sides of this condition shows that these equations
cannot be solved by finite-dimensional matrices. It is, however,
straightforward to solve them by self-adjoint operators on an
infinite-dimensional Hilbert space. The basic observation \cite{CDS} is that
consistency of the conditions (\ref{quotientconds}) when represented on this
Hilbert space requires that
\beq
\Omega_1\Omega_2=\ee^{2\pi i\theta}\,\Omega_2\Omega_1
\label{nc2torus}\eeq
for some real number $\theta$. This means that the operators $\Omega_\mu$
generate the noncommutative two-torus with noncommutativity parameter $\theta$.

The operators $\Omega_\mu$ may be represented on a Hilbert space ${\cal
H}_{p,q}$ of functions $f_k(s)$ where $s\in{\IR}$ and 
%%%%%%% final correction %%%%%%%
$k \in {\IZ}_q$
%%%% $1 \leq k\leq q$
\cite{CDS}. By introducing a fixed, fiducial derivation $\nabla_\mu$ on this
Hilbert space which satisfies
\beq
[\nabla_\mu,\Omega_\nu]=2\pi i\,\delta_{\mu\nu}\,\Omega_\nu \ ,
\label{nabladef}\eeq
a generic solution of (\ref{quotientconds}) may be taken to be the sum of
$\nabla_\mu$ and a fluctuating part,
\beq
X_\mu= iR_\nu\delta_{\mu\nu}\nabla_\nu+A_\mu(Z)
\label{Xgensoln}\eeq
where $Z_\mu$ generate the algebra of operators which commute with the
$\Omega_\mu$'s. There is a standard construction of these operators on the
Hilbert space ${\cal H}_{p,q}$ \cite{CDS}. In noncommutative geometry this
simply corresponds to the algebraic construction of vector bundles over the
noncommutative torus and the solutions (\ref{Xgensoln}) are just connections on
these bundles \cite{Connes}. The bundle ${\cal H}_{p,q}$ is characterized by
its ``commutative'' rank $p=\dim{\cal H}_{p,q}|_{\theta=0}$ and its magnetic
flux $q$ which is taken to be the constant curvature of the fixed connection
$\nabla_\mu$, $2\pi q={\rm Tr}\,i\,[\nabla_\mu,\nabla_\nu]$. The gauge fields
$A_\mu(Z)$ are then functions on a dual noncommutative torus and the
substitution of (\ref{Xgensoln}) back into the action (\ref{IKKT}) gives
Yang-Mills theory on this dual noncommutative torus.

This construction has been reinterpreted recently in terms of open string
quantization in the presence of a constant background $B$-field \cite{SW}. The
modules ${\cal H}_{p,q}$ are constructed from the boundary worldsheet theory
appropriate to one end of an open string terminating on a D2-brane and the
other end on a configuration of $p$ coincident D2-branes carrying $q$ units of
D0-brane charge. In this paper we will present a construction which is a
straightforward discretization of the above formalism in terms of an $N\times
N$ unitary matrix model. We shall recover all the parameters labeling the
continuum theory in the large $N$ limit. In particular, in this formulation
finite $\lambda$ is compatible with finite $L$. The proposal in \cite{AIIKKT}
can be regarded as a special case, from which it becomes transparent why finite
$\lambda$ is not compatible with finite $L$ in that instance. Thus the ensuing
matrix model naturally interpolates between the model in \cite{AIIKKT}
and the continuum formalism in \cite{CDS} for the
matrix model of M-theory \cite{BFSS}. We will show that our model can be
interpreted as a manifestly star-gauge invariant lattice formulation of
noncommutative gauge theory, which reduces to Wilson's lattice gauge theory
\cite{Wilson} for particular choices of the parameters even at finite $N$. We
shall also describe how various aspects of noncommutative gauge theory can be
systematically studied within the matrix model formalism.

\section{The Unitary Matrix Model}

\setcounter{equation}{0}

We will describe the construction in the simplest two-dimensional case,
but the generalization to arbitrary even dimension is straightforward.
The model we consider is just a twisted Eguchi-Kawai model \cite{EK,GO}, but
with a certain constraint imposed on the matrices. The action is
\beq
S= - \beta  \sum_{\mu \neq \nu} Z_{\mu\nu}\,
\tr\left( U_\mu U_\nu U_\mu ^\dag U_\nu ^\dag\right) 
\label{EKaction}
\eeq
where $U_{\mu}$ ($\mu=1,2$) are $N \times N$ unitary matrices and
$Z_{\mu\nu}=Z^*_{\nu\mu}$ is a phase factor called the ``twist''.
The constraint we impose on the matrices $U_\mu$ is
\beq
\Omega _\nu U_\mu \Omega_\nu ^\dag
= \ee ^{2 \pi  i \delta_{\mu\nu} r_\mu /N }\, U_\mu 
\label{constraint}
\eeq
where $r_\mu$ ($\mu = 1,2$) are integers which we will specify below. The
constraint \rref{constraint} is the exponentiation of Eq.~\rref{quotientconds}.
Now, however, the only condition required is that the $U_\mu$'s are traceless
unitary matrices. It therefore represents a finite dimensional version of the
quotient conditions for Matrix theory.

To solve the consistency conditions \rref{nc2torus}, we take the unitary
matrices
\beq
\Omega _1 = (\Gamma_2)^m  \otimes (\tilde{\Gamma} _1)^{\dag p}~~~~~~,~~~~~~
\Omega _2 =(\Gamma_1)^m  \otimes (\tilde{\Gamma}_2)^{\dag} \ ,
\eeq
where $\Gamma _\mu$ and $\tilde{\Gamma} _\mu$ are unitary matrices
of dimension $M$ and $q$, respectively, which satisfy the
Weyl-'t~Hooft commutation relations
\beq
\Gamma_1 \Gamma_2 = \ee ^{2 \pi i / M}\,\Gamma_2 \Gamma_1 ~~~~~~,~~~~~~
\tilde{\Gamma}_1 \tilde{\Gamma}_2
= \ee ^{2 \pi i / q}\,\tilde{\Gamma}_2 \tilde{\Gamma}_1 \ .
\label{GammaCR}
\eeq
These algebras can be represented by the shift and clock matrices
$(\Gamma_1)_{jk}=\delta_{j+1,k}$, $(\Gamma_2)_{jk}=(\ee^{2\pi
i/M})^{j-1}\,\delta_{jk}$, and similarly for the $\tilde\Gamma_\mu$. The
integers $M$ and $q$ satisfy $N=Mq$ and we take $M=mnq$. The
deformation parameter $\theta$ of Eq.~\rref{nc2torus} is given by
\beq
\theta =\frac{p}{q} - \frac{m}{nq} \ .
\eeq
The incorporation of two independent integers $m$ and $n$ in the above will
enable us to take a large-$N$ limit whereby the appropriate continuum
limit is reproduced.
They play a certain dual role to one another as we shall see. We will also see
later on that the meaning of the integers $p$ and $q$ will be the same as in
the description of the modules ${\cal H}_{p,q}$ above.

Our first task is to solve the constraint (\ref{constraint}).
We take the simplest particular solution $U_\mu=D_\mu $
associated with%
\footnote{\label{foot1} One can consider a more general particular
solution $U_\mu=(D_\mu)^{l_\mu}$ where $l_\mu$ are integers.
If $l_\mu$ is a divisor of $mq$, then this solution will
reproduce in the continuum limit
the noncommutative gauge theory associated
with a torus of modulus $R_1/R_2=l_1/l_2$.}
$r_1 = r_2 = mq $:
\beq
D_1 =(\Gamma _1)^\dag \otimes {\bf 1}_q ~~~~~~,~~~~~~
D_2 =\Gamma _2 \otimes {\bf 1}_q \ .
\eeq
These operators will become fixed covariant derivatives in the continuum limit.
We then decompose $U_\mu$ using $D_\mu $ as
\beq
U_\mu  = \tilde{U}_\mu D_\mu \  ,
\label{UvsU}
\eeq
where $\tilde{U}_\mu$ are unitary matrices which satisfy the constraint
\beq
\Omega_\nu \tilde{U}_\mu \Omega_\nu ^\dag = \tilde{U}_\mu \ .
\label{eq:tildeU}
\eeq
These constrained matrices will become the gauge fields of the model in the
continuum limit and they can be constructed
as follows. Assuming that $p$ and $q$ are co-prime, we choose integers $a$ and
$b$ such that
\beq
ap + bq = 1 \ .
\label{apbq}
\eeq
We then introduce unitary matrices
\beq
Z_1 =(\Gamma_2)^n  \otimes(\tilde{\Gamma} _1)^{\dag} ~~~~~~,~~~~~~
Z_2 =(\Gamma_1)^{\dag n}  \otimes(\tilde{\Gamma} _2)^a \ ,
\eeq
which commute with $\Omega_\mu$. The commutation relation of the $Z_\mu$ is
\beq
Z _1 Z _2 = \ee ^{2 \pi i \theta '} \,Z _2 Z _1 \ ,
\eeq
where $\theta '$ is given by
\beq
\theta ' = \frac{n}{mq} - \frac{a}{q}
\label{theta'}
\eeq
and it is related to $\theta$ through the discrete
M\"obius transformation
\beq
\theta ' = \frac{a \theta + b}{p- q \theta} \ .
\label{Mobius}
\eeq
In the continuum, the transformation law \rref{Mobius} would be just that
between Morita equivalent noncommutative tori \cite{Connes}. In fact, by
identifying $\theta$ with a constant Neveu-Schwarz two-form field, it is just
the T-duality transformation rule for the $B$-field \cite{SW,Morita}. The
relationship between Morita equivalence and duality \cite{Morita} means that
certain noncommutative gauge theories are physically equivalent to one another.
We will return to this point later on.

Using $Z_\mu$, we can define a basis for the solution space of
(\ref{eq:tildeU}) as
\beq
J_{m_1,m_2} = (Z_2)^{m_1} (Z_1)^{m_2}
{}~\ee ^{ \pi i \theta ' m_1 m_2 }
\label{defJ}
\eeq
where the phase factor is included so that
\beq
J_{-m_1,-m_2}=(J_{m_1,m_2}) ^ \dag \ .
\label{Jconjg}
\eeq
Since 
%%%%%%% final correction %%%%%%%
$(Z_\mu)^{mq} ={\bf 1}_N$, 
% $(Z_\mu)^{mq} = 1$, 
$J_{m_1,m_2}$ is periodic with respect to $m_1$ and
$m_2$ with period $mq$. We can therefore restrict the integers $m_1$ and $m_2$
to run from 0 to $mq-1$. 
It will prove convenient to introduce a lattice with $(mq)^2$ sites
on the torus and to instead work with the basis
defined by
\beq
\Delta (x) =  \sum _{m_1, m_2}
J_{m_1,m_2}~\ee ^{ - 2 \pi i  \epsilon_{\mu\nu} m_\mu x_\nu / L } \ ,
\label{defDelta}
\eeq
where $x_\mu= 0,\epsilon,\dots,\epsilon(mq-1)$ belongs to the lattice 
of the spacing $\epsilon$ and the 
extent of the lattice is 
\beq
L =  \epsilon mq.
\label{defL}
\eeq 
We have defined $\Delta (x)$ in such a way that the identities
\beqa
\frac{1}{(mq)^2}\sum _{x} \Delta (x) &=& {\bf 1}_N \ , \\
D _\mu  \Delta (x) D _\mu ^{\dag} &=& \Delta(x-\epsilon\hat{\mu})
\label{deltashift}
\eeqa
hold. Here $\epsilon\hat\mu$ denotes a shift by $\epsilon$ of $x_\mu$ only. 
The relation
\rref{deltashift} expresses the identification of the matrices $D_\mu$ as
discrete covariant derivatives. Note also that $\Delta (x)$ is hermitian, 
due to
(\ref{Jconjg}), and is periodic with respect to $x_1$ and $x_2$
with period $L$. The proof of completeness the generators \rref{defDelta} is
given in Appendix A. Given this complete set of solutions, we can write
a general solution to (\ref{eq:tildeU}) as
\beq
\tilde{U} _\mu  = \frac{1}{(mq)^2}\sum _{x} {\cal U}_\mu (x) \Delta (x) \ .
\label{mavsfu}
\eeq
Using orthogonality we can invert \rref{mavsfu} to give (see Appendix A)
\beq
{\cal U}_\mu(x)=\frac1N\,\tr\Bigl(\tilde U_\mu\,\Delta(x)\Bigr) \ .
\label{Uxinv}\eeq
In order that the right-hand side of Eq.~\rref{mavsfu} is a unitary matrix,
the coefficients ${\cal U} _\mu (x)$ should satisfy a certain
condition which will be given below.

Having solved the constraint, our next task is to rewrite our model entirely
in terms of ${\cal U} _\mu (x)$, which are the gauge fields
of the theory. For this, we use the identity \rref{Uxinv} to regard
$\Delta (x)$ as a map from the space of
$N \times N$ matrices which commute with $\Omega_\mu$
to the space of fields on a periodic $L \times L$ lattice.
We use the natural definition of a product of two lattice fields
$f_1(x)$ and $f_2 (x)$:
\beq
f_1 (x) \star f_2 (x) \defeq \frac 1N\,\tr\Bigl(f_1 f_2\,\Delta (x)\Bigr) \;,
\eeq
where $f_i$ are the $N\times N$ matrices defined
by $ f_i= (mq)^{-2}\sum_xf_i (x) \Delta (x)$.
This product is associative but not commutative.
One can write it explicitly in terms of $f_i (x)$ as
\beq
f_1 (x) \star f_2 (x) = \frac{1}{(mq)^2}\sum _{y,z}
f_1 (y) f_2 (z)~\ee
^{2 i B \epsilon_{\mu\nu} (x_\mu -y_\mu)(x_\nu - z_\nu)} 
\label{starprod}
\eeq
where 
\beq
B = \frac{2\pi}{\theta ' L^2} . 
\label{defB}
\eeq
These formulas are similar to Ref.~\cite{BM}.
(See also \cite{earlier} for earlier works in this regard.)
The product \rref{starprod} can be
considered as the lattice version of the star product
in noncommutative geometry. To see this, we note that in the continuum the star
product of two functions $f_1(x)$ and $f_2(x)$ may be defined as
\beq
f_1(x) \star f_2(x) \defeq f_1(x) ~
\exp \Bigl( i\frac 12 \overleftarrow{\del_\mu}\,\theta _{\mu\nu}
\overrightarrow{\del_\nu}\Bigr) ~f_2(x) \ .
\eeq
Using a Fourier transformation, this definition can be
turned into an integral form
\beq
f_1(x) \star f_2(x)  = \int\!\!\!
\int \dd y~\dd z ~K(x-y,x-z)\,f_1(y) f_2(z) 
\label{starcontinuum}
\eeq
where the integration kernel $K$ is given by
\beq
%%%%%%% final correction %%%%%%%
K(x-y,x-z) = \frac{1}{\pi^d |\det \theta_{\mu\nu}|}
%K(x-y,x-z) = \frac{1}{\pi^D |\det \theta_{\mu\nu}|}
\ee ^{ -2i (\theta ^{-1})_{\mu\nu}
(x_\mu -y_\mu) (x_\nu - z_\nu) } \ .
\eeq
The expression (\ref{starprod}) can be obtained
from (\ref{starcontinuum})
just by restricting the variables $x$, $y$, $z$ to run
over lattice points. In this sense, the product (\ref{starprod})
is a natural lattice counterpart of the star product
in the continuum. We shall therefore call
(\ref{starprod}) a star product in what follows.

Using the star product, we can write down the condition on
${\cal U}_\mu(x)$ which is required for $\tilde{U} _\mu$
to be unitary as
\beq
{\cal U} _\mu (x) ^ {\ast} \star {\cal U} _\mu (x) = 1 \ .
\eeq
In other words, the lattice fields ${\cal U}_\mu(x)$ must be star-unitary.
We may now rewrite the action (\ref{EKaction}) as
\beqa
S &=&  - \beta \frac{1}{(mq)^{2}} \sum _x  \sum_{\mu \neq \nu} Z_{\mu\nu}\,
\tr\left[ U_\mu U_\nu U_\mu ^\dag U_\nu ^\dag  \Delta (x)\right] \n
&=&  - \beta \frac{1}{(mq)^{2}} \sum _x  \sum_{\mu \neq \nu} Z_{\mu\nu}\,
\tr\left[ \tilde{U}_\mu D_\mu \tilde{U}_\nu D_\nu
D_\mu ^{ \dag} \tilde{U}_{\mu} ^\dag
D_\nu ^{ \dag} \tilde{U}_{\nu} ^\dag
\Delta (x)\right] \n
&=&  - \beta \frac{1}{(mq)^{2}} \sum _x  \sum_{\mu \neq \nu} 
\tilde{Z} _{\mu\nu}\,
\tr\left[ \tilde{U}_\mu  ( D_\mu  \tilde{U}_\nu D_\mu ^{\dag})
( D_\nu  \tilde{U}_\mu ^\dag D_\nu ^{\dag} )
\tilde{U}_\nu ^\dag \Delta (x)\right]  \n
&=& - \beta \frac{n}{m} \sum _x  \sum _{\mu \neq \nu} \tilde{Z}_{\mu\nu}~
{\cal U}_\mu (x) \star {\cal U}_\nu (x+\epsilon\hat{\mu}) \star
{\cal U}_\mu ^ \ast (x+\epsilon\hat{\nu}) \star
{\cal U}_\nu ^\ast (x) \ ,
\label{starwilson}
\eeqa
where $\tilde{Z}_{\mu\nu}= Z_{\mu\nu}\,\ee ^{-2 \pi i \epsilon_{\mu\nu}/M}$
can be considered as a background rank-two tensor field.
One can make $\tilde{Z}_{\mu\nu} =1$ by choosing
$Z_{\mu\nu} = \ee ^{2 \pi i \epsilon_{\mu\nu} /M}$.
Then the vacuum configuration is given by $\tilde U_\mu={\bf 1}_N$,
which corresponds to ${\cal U}_\mu (x) =1$,
up to the symmetry of the model which we now proceed to discuss.

The action (\ref{EKaction}) and the Haar integration
measure for the matrices $U_\mu$
are invariant under the SU($N$) transformations
\beq
U_\mu \rightarrow g\,U_\mu\,g^\dag \ .
\label{suN}
\eeq
The constraint (\ref{constraint}) in general breaks this symmetry
down to a subgroup of SU($N$).
However, the constrained model is still invariant under (\ref{suN})
for any $g$ that commutes with $\Omega _\mu$.
We can represent such a $g$ in terms of a function $g(x)$ on the lattice as
\beq
g   = \frac{1}{(mq)^2}\sum _{x} g(x) \Delta (x) \ ,
\eeq
where $g(x)$ should satisfy $
g (x) ^ {\ast} \star g (x) = 1$, but is otherwise arbitrary.
The transformation (\ref{suN}) can now be interpreted in
terms of ${\cal U} _\mu (x)$ as
\beq
{\cal U}  _\mu(x)  \rightarrow
g(x) \star {\cal U} _\mu  (x) \star g^\ast (x+\epsilon\hat{\mu}) \ .
\label{onelink}
\eeq
Therefore, the resulting theory of the lattice field
${\cal U} _\mu (x)$ is manifestly invariant under
this star-gauge symmetry.

We can show that the theory \rref{starwilson} reduces to Wilson's
lattice gauge theory \cite{Wilson} for particular choices of the parameters.
Note that we can always make $\theta ' = 0$ by taking $n = m a$.
In this case, the star product becomes the ordinary product of functions.
Therefore, ${\cal U}_\mu (x)$ becomes an ordinary U(1) field
on the lattice and the action (\ref{starwilson}) becomes the ordinary
Wilson plaquette action. We can also show that the integration
measure for ${\cal U}_\mu (x)$ is actually the Haar measure
for integration over the group U(1)$^{2 (mq)^2}$.
Note that the Haar measure for the $N \times N$ matrices $U_\mu$
and the constraint (\ref{constraint}) are invariant under
$U_\mu \rightarrow g U_\mu$ for any $g$ which commutes with $\Omega_\mu$.
This can be translated into the invariance of the integration measure
for ${\cal U}_\mu (x)$ under ${\cal U}_\mu (x) \rightarrow g(x) {\cal U}_\mu
(x)$. The uniqueness of a measure with such an invariance proves
our statement. Thus, our lattice formulation of noncommutative gauge theory
includes Wilson's lattice gauge theory on a periodic lattice of finite extent
as the $\theta ' = 0$ case, even at finite $N$. We remark that in this case,
although the $Z_\mu$ matrices can be taken to be diagonal, 
the $(mq)^2$ degrees of freedom of the lattice gauge theory
are contained in the $N=mnq^2=a(mq)^2$ diagonal elements
of $\tilde U_\mu$.

%%%%%%% final correction %%%%%%%
Going back to the general case of arbitrary $\theta '$,
let us now consider 
%%%%Let us now consider 
the continuum limit of the model \rref{starwilson}
when the lattice spacing $\epsilon\rightarrow 0$. 
We introduce the continuum field 
$\tilde
A_\mu$ and operator $d_\mu$ through
\beq
\tilde{U}_\mu = \ee ^{i \epsilon \tilde{A}_\mu}~~~~ , ~~~~
D_\mu = \ee^{i \epsilon d_\mu} \ .
\label{continuum}
\eeq
The large $N$ limit dictated by the continuum theory \cite{CDS}
is $m \sim n \sim \sqrt{N}$ and $\epsilon \sim 1/\sqrt{N}$ with fixed $a$, $b$,
$p$ and $q$. Both $L$ given by~\rref{defL} and ${B}$ given by~\rref{defB}
are finite in such a large $N$ limit.
The resulting gauge theory is constructed from connections of a rank $p$ bundle
of magnetic flux $q$. We will see this explicitly in
the next section. Note that the field
theory \rref{starwilson} is actually of rank 1. This is one of the
characteristic features of Morita equivalence or alternatively of T-duality
transformations between different brane worldvolume field theories. The
original SU($p$) Yang-Mills theory on the noncommutative torus with deformation
parameter $\theta$ is physically equivalent to a U(1) Yang-Mills theory on a
dual torus with noncommutativity parameter (\ref{Mobius}) that implicitly
contains the information about the rank $p$ of the underlying vector bundle.
The case $q=0$, representing a trivial gauge bundle, can also be constructed
and will be presented elsewhere.

However, as far as the continuum limit of the lattice theory
is concerned, we need only send $m$ to infinity, but not necessarily $n$.
If $n$ is finite as $m\rightarrow\infty$,
this does not lead to the solutions constructed in the continuum
\cite{CDS} for hermitian operators and is instead
associated with unitary operators acting on periodic functions
of $0\leq s < nq$. The particular case of
$q=n=1$, for which the condition (\ref{constraint}) is trivial
and our model reduces to the ordinary, unconstrained
twisted Eguchi-Kawai model, is of this type. It corresponds to
the interpretation of the twisted large $N$ reduced model
in terms of noncommutative gauge theory which was proposed in
\cite{AIIKKT}. Since $\theta ' = 1/N$ in that case,
in order to make ${B}$ finite one needs $\epsilon \sim 1/\sqrt{N}$,
which inevitably makes the physical extent of the torus
scale as $L \sim \sqrt{N}$, reproducing the observation made in \cite{IIKK}.
Note that the issue of whether or not a continuum limit really exists is a
dynamical question that can be addressed, for example, by Monte Carlo
simulation.
A numerical simulation of the two-dimensional Eguchi-Kawai model
has been done in \cite{NN}, where a non-trivial large $N$
scaling behavior was found with the parameter $N \epsilon ^2$ fixed,
which is exactly the large $N$ limit required to make the
physical scale of noncommutativity finite.
This in itself means that noncommutative gauge theory with
a background tensor field can be constructively defined.

\section{Observables of Noncommutative Gauge Theory}

\setcounter{equation}{0}

We will now describe how the properties of noncommutative gauge theory can be
completely reformulated in the language of the unitary matrix model above.
Let us define a lattice path which consists of $\K$ links by
$C=\{\hat{\mu}_1, \ldots,\hat{\mu}_\K \}$
and $C^{-1}=\{\hat{\mu}_\K, \ldots,\hat{\mu}_1 \}$ for
an opposite orientation.
The path $C$ connects lattice sites separated by the vector
$\vl^\mu=\xi^\mu_\K$ while
\beq
\xi^\mu_i= \epsilon \sum\limits_{j=1}^i \hat{\mu}_j 
\label{xi}
\eeq
belongs to $C$.
We introduce the following products of matrices along the path:
\beqa
D(C)&=&\prod\limits_{j=1}^\K D_{\mu_j}~~~~~~,~~~~~~
D(C^{-1})=D(C)^\dagger \ ; \n
U(C) &=&\prod\limits_{j=1}^{\K}
\left(\tilde U_{\mu_j}D_{\mu_j}\right)  \ .
\label{mproducts}
\eeqa
Given the property~\rref{deltashift} we then have
\beq
\Delta(x+\vl)=D(C)\Delta(x) D(C^{-1}) \ ,
\eeq
where the right-hand side is path-independent because of the properties
of the $D_{\mu}$.
This results in the formula
\beq
\frac 1N\,\tr \l A\,\Delta(x)\r \star \frac 1N\,
\tr \l B\,\Delta(x+\vl)\r
=\frac 1N\,\tr\l A D(C) B D(C^{-1})\,\Delta(x)\r \ ,
\label{AstarB}
\eeq
provided that $A$ and
$B$ belong to the commutant of the algebra generated by $\Omega_\mu$.
Using \rref{AstarB}, we can construct the matrix analog of
the noncommutative phase factor along the lattice path which defines parallel
transport for the gauge bundle in the continuum limit,
\beq
{\cal U}(x;C)\defeq  
\star\prod\limits_{j=1}^\K {\cal U}_{\mu_j}(x+\xi_{j-1})
= \frac 1N\,\tr \l U(C) D(C^{-1}) \,\Delta(x)\r 
\label{U(C)}
\eeq
where the product in the middle is the star product.
Under the SU($N$) gauge transformation \rref{suN} where
$U(C)\rightarrow g  U(C) g^\dagger $,
the right-hand side of Eq.~\rref{U(C)} transforms as
\beq
{\cal U}(x;C) \rightarrow
\frac 1N\,\tr \l g {U}(C) g^\dag D(C^{-1})\, \Delta(x)\r=
g(x)\star {\cal U}(x;C)\star g^\ast (x+\vl)
\label{gaugecovariant}
\eeq
as it should for the phase factor. This formula extends \rref{onelink}
to an arbitrary open path.

The continuum limit of the above construction is given
by the large-$N$ limit of the matrix model. We introduce
$\dd\xi^\mu=\epsilon \hat\mu$, so that Eq.~\rref{xi} takes the form
\beq
\xi^\mu=\int\dd\xi^\mu \ ,
\label{cxi}
\eeq
and write down the continuum analogs of Eqs.~\rref{mproducts} and \rref{U(C)}
using \rref{continuum} as
\beqa
D(C) &= & {\rm P}\, \exp \l{i\int_0^\vl\dd\xi^\mu~d_\mu}\r \ ,\n
 U(C)& =& {\rm P}\,
\exp\l {i\int_0^\vl\dd\xi^\mu~(\tilde A_\mu +d_\mu ) }\r
\label{cmproduct}
\eeqa
and
\beq
{\cal U}(x;C) =  \star\prod\limits_\xi
\l 1+i\,\dd\xi^\mu\,{\cal A}_{\mu}(x+\xi)\r .
\label{cU(C)}
\eeq
Here we have defined the field ${\cal A}(x)$ by
\beq
{\cal U} _\mu (x)=\star\,\ee^{i\epsilon {\cal A}_\mu (x)}
\label{stare}
\eeq
where the exponential is understood as a power series expansion
with the star-product.
Notice that the $d_\mu$  in Eq.~\rref{cmproduct} cannot be absorbed
by a shift of $\tilde A_\mu$ since $d_\mu$ does not commute with the
$\Omega$'s. This is the difference between the present model and
 the continuum limit of the twisted Eguchi-Kawai model where this translation
is usually done. 

The phase factors \rref{cU(C)} can be used to define a new class
of observables in the matrix model, associated with noncommutative
gauge theory. The standard closed Wilson loops $W(C)$ of twisted
reduced models~\cite{GO} which are invariant under~\rref{suN}
can be expressed via ${\cal U}(x;C)$ as
\beqa
W(C)&\equiv& \frac 1N\,\tr D(C^{-1})\;
\frac 1N\,\tr  U(C)  \n &=&
\frac{1}{(mq)^2}\sum_x \frac 1N\,\tr \l  U(C)
D(C^{-1}) \,\Delta(x)\r =\frac{1}{(mq)^2}\sum_x {\cal U}(x;C) \ ,
\label{Wstandard}
\eeqa
since $D(C^{-1})$ is a c-number. 
Therefore, the analog of $W(C)$
in noncommutative gauge theory is a sum over lattice points
of ${\cal U}(x;C)$,
which is understood as the sum over translations of the closed path
that preserve its shape. This object is star-gauge invariant
due to this additional summation.
For the simplest closed loop, i.e. the plaquette, it is used in
constructing the action~\rref{EKaction}.
What is rather surprising in noncommutative gauge theory is that
one can actually construct a
star-gauge invariant observable associated with
an open path, as has been found in~\cite{IIKK}.
We will now describe how such observables appear in our model
and point out an interesting consequence of the finiteness
of the spacetime extent.

Star-gauge invariant quantities can be constructed 
out of (\ref{U(C)}) with the aid of a lattice function $S_\vl(x)$ 
which has the property
\beq
S_\vl(x) \star g(x) \star S_\vl (x)^{-1} = g(x+\vl) 
\label{Svprop}\eeq
for arbitrary star-unitary functions $g(x)$. Here again
$\vl_\mu$ is the relative separation vector between
the two ends of the open loop. 
Star-gauge invariant quantities  can then be defined 
by $(mq)^{-2}\sum_x S_\vl(x)\star {\cal U}(x;C)$.
The property \rref{Svprop} in the matrix model becomes
\beq
S_\vl \Delta (x) S_\vl ^{-1} = \Delta (x-\vl) 
\eeq
where 
\beq
S_\vl = \frac{1}{(mq)^2}\sum _x  S_\vl (x) \Delta (x)
\eeq 
belongs to the commutant 
of the algebra generated by $\Omega_\mu$.
Using the definition (\ref{defDelta}) and with a little algebra,
we obtain that $S_\vl(x)$ should satisfy
\beq
S_\vl (x +  \theta '  L \hat{\mu}) 
= \ee ^{- 2 \pi i \epsilon _{\mu\nu} \vl_\nu /L }
S_\vl (x) 
\label{Svequ}
\eeq
where $L$ is given by~\rref{defL}.
Assuming that $ \theta ' mq = n - ma$ and $mq$ are co-prime,
the only solution to (\ref{Svequ}) is
\beq
S_\vl (x) = \ee ^{2 \pi i k \cdot x /L }
\label{Sv}
\eeq
where $k_\mu = 0,1,\cdots,(mq-1)$ and
\beq 
\vl_\mu = \theta ' L \epsilon _{\mu \nu} k _\nu  + n_\mu L 
\label{distance}
\eeq
with an integer vector $n_\mu$. As is seen from~\rref{Sv}, the ratio
$2 \pi k_\mu / L $ plays the role of the  momentum variable and it is
 related to the distance vector $\vl_\mu$ by Eq.~\rref{distance}.
The longer the open loop is, the larger the momentum $2 \pi k_\mu / L $
should be. 
The discretization of momentum due to the finite extent of 
the torus leads us to an interesting consequence that 
${\vl}_\mu$ should also be discrete. 
In the commutative case when $\theta ' = 0$, we obtain 
${\vl}_\mu = n_\mu L$  
reproducing the known fact that the only such 
gauge invariant quantities are the Polyakov loops (holonomies
of noncontractable loops on the torus).
It is remarkable that 
in noncommutative gauge theory on a finite volume there exist
 other objects of this kind with 
discretized values of the distance ${\vl}_\mu$.
It remains discrete in the continuum limit since $L$ is finite.
This is the difference from the analogous quantities
constructed in~\cite{IIKK} for the IIB model, where
${\vl}_\mu$ can be an arbitrary vector in the large $N$ limit. 
The matrix description of the star-gauge invariant open loop
is given by
\beqa
\frac{1}{(mq)^2}\sum_x S_\vl(x)\star {\cal U}(x;C) 
&=& \frac{1}{(mq)^2}\sum _x \frac 1N\,
\tr \l U(C) D(C^{-1} )S_\vl \,\Delta(x)\r \ \n
&=& \frac 1N\,\tr \l U(C) D(C^{-1} )S_\vl \r 
\label{opendef}
\eeqa
where $
S_\vl=J_{k_2,-k_1}$ is given by~\rref{defJ} for the solution~\rref{Sv}.
Its star-gauge invariance can be directly
checked by noting that $D(C^{-1} )S_\vl$ in (\ref{opendef})
belongs for $\vl$ given by Eq.~\rref{distance}
to the commutant of the algebra generated by $Z_\mu$, i.e.\
commutes with $g$.

The matrix model determines the dynamics of noncommutative gauge theory.
Let us demonstrate how the classical equation of motion emerges
in the matrix language. For simplicity we take the continuum limit
using the relation~\rref{continuum}. The continuum action
reads
\beq
S[\tilde A]= \tr \l\left(\tilde F_{\mu\nu} - f_{\mu\nu} \right)^2 \r
+\tr \l \alpha_{\mu\nu}\left[\tilde A_\mu, \Omega_\nu\right]\r
\label{caction}
\eeq
where
\beq
\tilde F_{\mu\nu}=id_\mu \tilde A_\nu -id_\nu \tilde A_\mu
+i\left[\tilde A_\mu,\tilde A_\nu\right] \ .
\label{defF}
\eeq
Here $f_{\mu\nu}$ is the constant curvature of the gauge bundle given by
\beq
-i\, [d_\mu,d_\nu] =  f_{\mu\nu}\, {\bf 1}_N 
\eeq
where in the two dimensional case
\beq
f_{\mu\nu}=\frac{2\pi q}{p-q\theta}\,R_1R_2\,\epsilon_{\mu\nu} \ .
\label{deff}\eeq
In the construction of section 2, $R_1=R_2=1/\epsilon nq$ are the radii of the
two-torus (see\ footnote~\ref{foot1}). 
Eq.~\rref{deff} is the standard formula for the curvature of the
module ${\cal H}_{p,q}$ \cite{CDS}. It should be understood, however, as being
multiplied by the identity operator ${\bf1}_{p,q}$ with ${\rm
Tr}\,{\bf1}_{p,q}=p-q\theta$, so that the integral curvature of the bundle is
%%%%%%% final correction %%%%%%%
${\rm Tr}\,f_{\mu\nu}/(2 \pi R_1 R_2) = q$. 
%%%%${\rm Tr}\,f_{\mu\nu}/R_1R_2=2\pi q$. 
In the present case this trace operation
corresponds to multiplying the curvature \rref{deff} by the dimensionless area
factor 
%%%%%%% final correction %%%%%%%
$\sqrt{(R_1R_2)(L_1L_2)}=m/n$ 
%%%%$(R_1R_2)(L_1L_2)=m^2/n^2$ 
giving the volume of a unit cell in the ``phase
space'' of the $d_\mu$'s. This is analogous to the derivation of the dimension
of the Hilbert space ${\cal H}_{p,q}$ presented in \cite{CDS}. The (infinite)
hermitian matrices in \rref{caction} are unconstrained while the constraints
are taken into account by the Lagrange multipliers $\alpha_{\mu\nu}$.
The action~\rref{caction} is of the type considered in~\cite{GAKA}, but
now with the additional constraints imposed on $\tilde A$.

The variational derivative
\beq
\frac{\delta}{\delta {\cal A}_\mu (x)}\,{\cal A}_\nu (y) =
\delta_{\mu\nu}\,\delta(x-y)
\label{delAA}
\eeq
can be represented in the matrix language as follows.
Given~\rref{mavsfu}, \rref{continuum} and \rref{stare}, we have
\beq
\tilde A_\mu = \int\dd x~{\cal A}(x) \Delta(x)
\eeq
which implies
\beq
\frac{\delta}{\delta {\cal A}_\mu (x)} =
\tr \l\frac {\partial}{\partial \tilde A_\mu}\,\Delta (x)\r \ .
\label{varder}
\eeq
Equation~\rref{delAA} is now reproduced as
\beq
\tr \left(\frac {\partial}{\partial \tilde A_\mu}\,\Delta (x) \right)
{}~\frac 1N\,\tr \l \tilde A_\nu\,\Delta (y)\r =
\delta_{\mu\nu}\,\frac 1N\,\tr \l\Delta(x) \Delta(y)\r
= \delta_{\mu\nu}\,\delta(x-y)
\label{asshould}
\eeq
as it should. We can treat the matrix elements of $\tilde A_\nu$ as
independent because of the completeness of the generators of the commutant.
Notice that there is an ordinary product of the traces
in \rref{asshould} rather than the star product.
Acting by \rref{varder} on the action \rref{caction}, we get
\beqa
\frac{\delta}{\delta {\cal A}_\nu (x)}\,S &=&
\tr \l\left[d_\mu+ \tilde A_\mu, \tilde F_{\mu\nu} - f_{\mu\nu}
\right]\,\Delta (x)\r +
\tr \l\alpha_{\mu\nu}\,\left[\Delta(x), \Omega_\mu\right] \r \n &=&
\tr \l\left[d_\mu+ \tilde A_\mu, \tilde F_{\mu\nu}\right]\,\Delta (x)\r \ .
\label{maxwell}
\eeqa
Eq.~\rref{maxwell} reproduces the noncommutative Maxwell equation.

The matrix representation~\rref{varder} of the variational
derivative is actually most useful for deriving the Schwinger-Dyson
equations of the quantum noncommutative theory and, in particular, the
loop equations.
To illustrate the technique, let us first calculate how the variation
${\delta}/{\delta {\cal A}_\mu (z)}$ acts on the noncommutative
phase factor ${\cal U}(x;C)$, which determines the contact
term in the loop equation~\cite{MM79}. Using~\rref{varder}, we get
\beqa
\frac{\delta}{\delta {\cal A}_\nu (z)}\,{\cal U}(x;C)&=&
\tr \left(\frac {\partial}{\partial \tilde A_\nu}\,\Delta (z) \right)~
\frac 1N\,\tr \l  U(C) D(C^{-1})\,\Delta(x)\r \n
&= &i \int_0^\vl\dd\xi^\nu~\frac 1N\,\tr \l  U(C_1)\,\Delta(z)
 U(C_2)\,D(C^{-1}) \Delta(x)\r \n
&= & i\int_0^\vl\dd\xi^\nu~{\cal U}(x;C_1)\star \delta(x+\xi-z) \star
{\cal U}(x+\xi; C_2)
\label{RHS}
\eeqa
where $C_1$ and  $C_2$ are the parts of the contour
$C$, $C=C_1C_2$, separated by $\xi$.
We can similarly calculate how the area operator
$\partial_\mu \delta /\delta \sigma^{\mu\nu}(z)$ ($z\in C$)
acts on ${\cal U}(x;C)$. This calculation is purely geometrical and gives
\beqa
\lefteqn{
\partial_\mu\,\frac{\delta }{\delta\sigma^{\mu\nu}(z)}\,
{\cal U}(x;C)=\frac 1N\,\tr \l  U(C_1)
\left[d_\mu+ \tilde A_\mu, \tilde F_{\mu\nu}\right] U(C_2)
 D(C^{-1})\,\Delta(x)\r }\n
&=& -i\;{\cal U}(x;C_1)\star \Bigl(
 \partial_\mu {\cal F}_{\mu\nu} +i {\cal A}_\mu\star
{\cal F}_{\mu\nu}-i{\cal F}_{\mu\nu}\star{\cal A}_\mu
\Bigr)(z)\star{\cal U}(z;C_2) 
\label{LHS}
\eeqa
where 
\beq
{\cal F}_{\mu\nu}=\partial_\mu{\cal A}_\nu-\partial_\nu{\cal A}_\mu
+i{\cal A}_\mu\star{\cal A}_\nu-i{\cal A}_\nu\star{\cal A}_\mu.
\eeq
That is, the operator $\partial_\mu \delta /\delta \sigma^{\mu\nu}(z)$
inserts the Maxwell equation in the noncommutative phase factor at the
point $z$, as anticipated.

The standard loop equation of large-$N$ Yang-Mills theory for
the Wilson loop average
$\langle W(C) \rangle$ emerges from Eqs.~\rref{RHS} and \rref{LHS} 
in the $q=n=1$ case by putting $z=x$, summing over all $x$
and using the formula
\beq
\frac{1}{N^2}\sum_x \Delta^{ij}(x)\Delta^{kl}(x) = 
N\,\delta^{il} \delta^{kj}~~~~(q=n=1~~{\rm case}) \ .
\eeq
This equation is quadratic in $\langle W(C) \rangle$ due to
large-$N$ factorization of correlators.
To better understand the
consequences of Eqs.~\rref{RHS} and \rref{LHS}
for $q,n\neq 1$, let us consider the case of
$\theta'=0$, whereby the continuum limit is rank one commutative gauge theory,
previously known as Maxwell's theory. The phase factor~\rref{cU(C)} for a
closed loop is now gauge invariant since the star-product becomes the ordinary
product, so that ${\cal U}(x;C)$ becomes the usual phase factor of
electrodynamics which is independent of $x$ while the
$g$'s cancel on the right-hand side of Eq.~\rref{gaugecovariant}. The loop
equation for the average of the phase factor
can be obtained by combining the averages of Eqs.~\rref{RHS} and \rref{LHS}
which results in the standard linear loop equation
\beq
\partial_\mu\,\frac{\delta }{\delta\sigma^{\mu\nu}(z)}\,
\Bigl\langle{\cal U}(C)\Bigr\rangle = \frac{1}{\beta \epsilon^4}
\int\dd\xi^\nu~\delta(\xi-z)\,\Bigl\langle{\cal U}(C)\Bigr\rangle \ .
\eeq
We have just illustrated by this simple example
how the phase factors~\rref{cU(C)} can
indeed correspond to observables in noncommutative gauge
theories associated with the unitary matrix model. 
This is precisely the novel feature of the present matrix
model that was pointed out in section 2, namely that in the large $N$ limit it
is possible to arrive at a U(1) continuum gauge theory.

\subsection*{Acknowledgements}

We would like to thank S. Iso for helpful correspondence related
to his work. The work of J.A. and Y.M. is supported in part by
MaPhySto founded by the Danish National Research Foundation. 
The work of Y.M. is supported in part by the grant RFFI
97--02--17927. J.N. is supported by the Japan Society for the Promotion of
Science as a Research Fellow Abroad. The work of R.J.S. is supported in part by
the Danish Natural Science Research Council.

\renewcommand{\theequation}{\thesection.\arabic{equation}}
\setcounter{section}{0}

\appendix{Proof of Completeness}

We will demonstrate that the $\Delta (x)$ defined by Eq.~\rref{defDelta}
form a complete set for the space of solutions to the constraints
(\ref{eq:tildeU}), i.e. that any $N \times N$ complex matrix $A$
that commutes with $\Omega_\mu$ ($\mu = 1,2$) can be written uniquely as
\beq
A = \frac{1}{(mq)^2}\sum _x A(x) \Delta (x) \ .
\label{AtoAx}
\eeq
We first note that
\beq
\E \defeq \Bigl\{ A \in \mbox{gl}(N,\IC) ~\Bigm| ~
A \Omega_\mu =  \Omega_\mu A~~,~~\mu=1,2\Bigr\}
\eeq
defines a linear subspace of gl($N$,$\IC$) which
has an inner product
defined by $\tr (A^\dag B)$ for $A,B \in \mbox{gl}(N,\IC)$.
The $\Delta (x) \in \E$ form an 
%%%%%%% final correction %%%%%%%
orthogonal
%%%%%orthonormal 
set,
\beq
\frac 1N\,\tr\Bigl(\Delta (x) \Delta (y)\Bigr) = 
(mq)^2 \delta _{x,y} \ .
\label{orthogonal}
\eeq
We now consider the linear subspace $\E'$ of $\E$ spanned by $\Delta (x)$.
We wish to show that $\E'=\E$. For this, we introduce a convenient
%%%%%%% final correction %%%%%%%
orthogonal
%%%%%orthonormal 
basis of gl($N$,$\IC$). Define $\tilde{\Delta} (\tilde{x})$ by
\beq
\tilde{\Delta} (\tilde x) =  \sum _{m_1, m_2}
(\Omega_1)^{m_1} (\Omega_2 )^{m_2} ~
\ee ^{ -\pi i \theta  m_1 m_2 } ~
\ee ^{  2 \pi i  \theta \epsilon_{\mu\nu} m_\mu\tilde x_\nu } \ ,
\label{defDeltatilde}
\eeq
where $\tilde{x}$ runs from 0 to $nq -1$
and we put $\epsilon=1$ for simplicity.
These matrices commute with $Z_\mu$, they are mutually 
%%%%%%% final correction %%%%%%%
orthogonal,
%%%%%orthonormal, 
and they satisfy
$\Omega_\mu \tilde{\Delta} (\tilde{x}) \Omega_\mu ^\dag =
\tilde{\Delta} (\tilde{x}- \hat{\mu}) $.
We take $\Delta (x) \tilde{\Delta} (\tilde{x})$
as an 
%%%%%%% final correction %%%%%%%
orthogonal
%%%%%orthonormal 
basis of gl($N$,$\IC$).
We now consider a generic element which belongs to
the orthogonal complement of $\E'$ in gl$(N,\IC)$ given by
\beq
\sum_{x,\tilde{x}\ne 0} f(x,\tilde{x})\,
\Delta (x) \tilde{\Delta} (\tilde{x}) \ .
\eeq
Requiring that it commutes
with both $\Omega_1$ and $\Omega_2$ implies immediately that
$f(x,\tilde{x}) \equiv 0$, which completes the proof.
Using the orthogonality (\ref{orthogonal})
of the basis $\Delta (x)$, we can write the $A(x)$
in (\ref{AtoAx}) as
\beq
A(x) = \frac 1N\,\tr\Bigl(A\,\Delta (x)\Bigr) \ .
\eeq

\end{document}